\documentclass[letterpaper]{article}
\usepackage[table]{xcolor}
\usepackage[utf8]{inputenc}
\usepackage{aaai25}
\usepackage{times}
\usepackage{helvet}
\usepackage{courier}
\usepackage[hyphens]{url}
\usepackage{graphicx}
\usepackage{natbib}
\usepackage{caption}
\usepackage{booktabs}
\usepackage{placeins}
\usepackage{textgreek}
\usepackage{tabularx}
\usepackage{algorithm}
\usepackage{algorithmic}
\usepackage{newfloat}
\usepackage{listings}

\usepackage{tikz}
\usepackage{pgfplots}
\usepackage{pgfplotstable}
\pgfplotsset{compat=1.18}
\usepackage{amsmath}
\usepackage{comment}

\DeclareCaptionStyle{ruled}{labelfont=normalfont,labelsep=colon,strut=off}
\floatstyle{ruled}
\newfloat{listing}{tb}{lst}{}
\floatname{listing}{Listing}

\title{Understanding Privacy Norms Around LLM-Based Chatbots: \\A Contextual Integrity Perspective}

\author {
    Sarah Tran\textsuperscript{\rm 1},
    Hongfan Lu\textsuperscript{\rm 1}, 
    Isaac Slaughter\textsuperscript{\rm 1}, 
    Bernease Herman\textsuperscript{\rm 1}, 
    Aayushi Dangol\textsuperscript{\rm 1}, 
    Yue Fu\textsuperscript{\rm 1}, 
    Lufei Chen\textsuperscript{\rm 1}, 
    Biniyam Gebreyohannes\textsuperscript{\rm 1}, 
    Bill Howe\textsuperscript{\rm 1}, 
    Alexis Hiniker\textsuperscript{\rm 1},
    Nicholas Weber\textsuperscript{\rm 1}\equalcontrib,
    Robert Wolfe\textsuperscript{\rm 2}\equalcontrib\thanks{Work performed while at the University of Washington}
}
\affiliations {
    \textsuperscript{\rm 1}University of Washington\\
    \textsuperscript{\rm 2}Rutgers University\\
    saraht45[at]uw.edu,
    nmweber[at]uw.edu,
    robert.wolfe[at]rutgers.edu
}

\begin{document}

\maketitle

\begin{abstract}
LLM-driven chatbots like ChatGPT have created large volumes of conversational data, but little is known about how user privacy expectations are evolving with this technology. We conduct a survey experiment with 300 US ChatGPT users to understand emerging privacy norms for sharing chatbot data. Our findings reveal a stark disconnect between user concerns and behavior: 82\% of respondents rated chatbot conversations as sensitive or highly sensitive — more than email or social media posts — but nearly half reported discussing health topics and over one-third discussed personal finances with ChatGPT. Participants expressed strong privacy concerns (t(299) = 8.5, p $<$.01) and doubted their conversations would remain private (t(299) = -6.9, p  $<$.01). Despite this, respondents uniformly rejected sharing personal data (search history, emails, device access) for improved services, even in exchange for premium features worth \$200. To identify which factors influence appropriate chatbot data sharing, we presented participants with factorial vignettes manipulating seven contextual factors. Linear mixed models revealed that only the transmission factors such as informed consent, data anonymization, or the removal of personally identifiable information, significantly affected perceptions of appropriateness and concern for data access. Surprisingly, contextual factors including the recipient of the data (hospital vs. tech company), purpose (research vs. advertising), type of content, and geographic location did not show significant effects. Our results suggest that users apply consistent baseline privacy expectations to chatbot data, prioritizing procedural safeguards over recipient trustworthiness. This has important implications for emerging agentic AI systems that assume user willingness to integrate personal data across platforms.
\end{abstract}
\section{Introduction}

\begin{figure*}
    \centering
    \includegraphics[width=.95\linewidth]{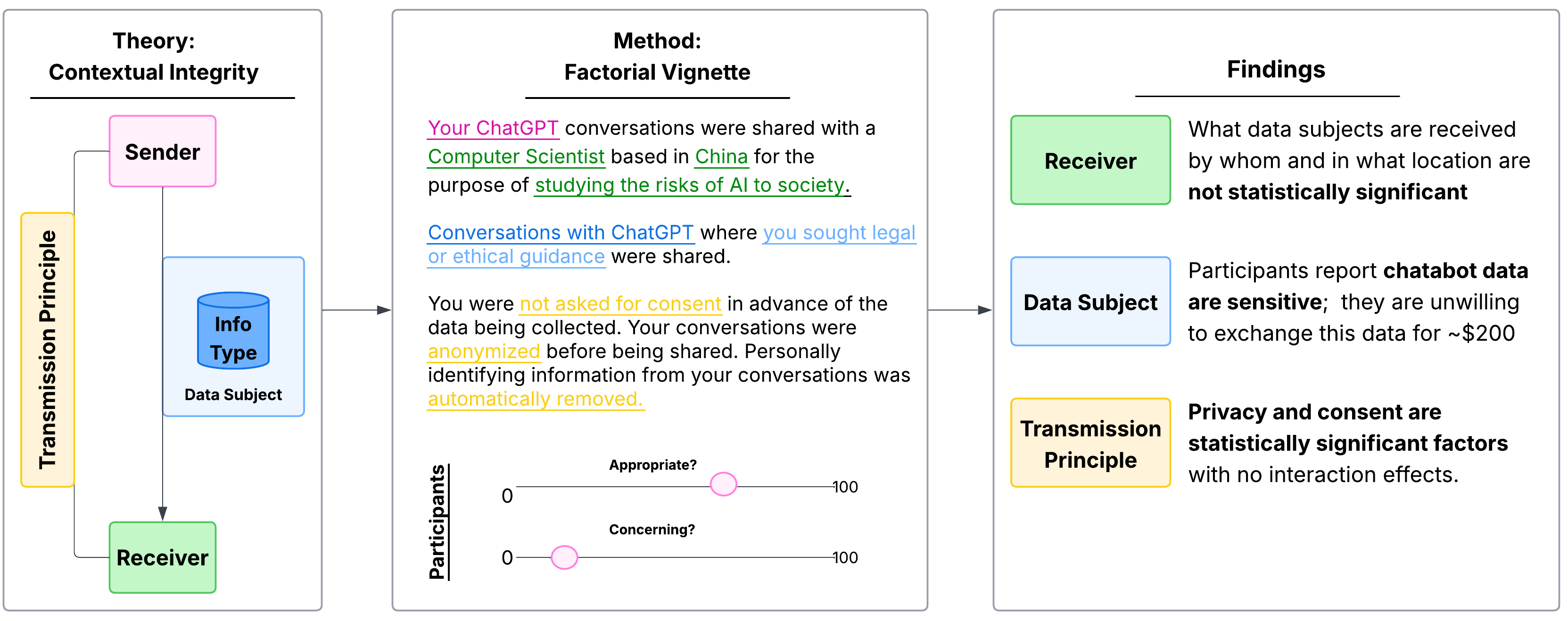}
    \caption{We used the contextual integrity framework to construct factorial vignettes rated by participants based on whether they were appropriate or concerning. This approach allowed us to isolate factors associated with the \textbf{transmission principle} - the set of norms and expectations surrounding data exchange - as the primary determinants of privacy norms for chatbot data.}
    \label{fig:teaser}
\end{figure*}

Conversational AI systems like ChatGPT have rapidly become a source of user-generated data, with over 400 million weekly active users creating billions of chat interactions \cite{chatgptusercount2025}. This conversational data represents a fundamentally new category of personal information — more intimate than search queries, more extensive than social media posts, and often containing highly sensitive disclosures about health, finances, and personal relationships. The commercial and strategic value of chatlog data spans targeted advertising \cite{matz2024potential}, personalized interfaces \cite{chen2024large}, training next-generation language models \cite{nasr2023scalable,nasr2025scalable}, and even geopolitical leverage when nations control chat data from other countries' citizens \cite{deepseekshock2025,wired2025deepseek}.
Yet users remain uncertain about how their conversational data is handled. While chatbot providers may offer privacy protections through terms of service agreements, research indicates most users neither understand these policies nor trust that their data will be used responsibly. A 2023 Pew Research Center study found that large majorities of Americans familiar with AI worry about personal information being used in ``unintended ways and ways people are not comfortable with'' \cite{pew2023aiprivacy}. These concerns have proved prescient: In December 2024, the chatbot provider WotNot inadvertently exposed over 340,000 private chat logs through an unprotected cloud system \cite{malwarebytes2024chatbot}, and in August 2025, researchers discovered that Google was indexing and making discoverable ChatGPT conversations that users had inadvertently shared - exposing passport numbers, medical records, and employment histories \cite{Cox_2025}. 

Despite the apparent risks, prior research finds that users of AI chatbots often disclose sensitive information in chat logs, sometimes relying on flawed mental models of how data will be processed and retained by a model's provider \cite{zhang2024s}. Indeed, the presence of user social security numbers and other personally identifiable information (PII) resulted in the withdrawal of ShareGPT, an open dataset of interactions with ChatGPT used in prior privacy studies \cite{zhang2024s}. The apparent gap between many users' behavior when interacting with chatbots (\textit{e.g.,} sharing sensitive PII) and the public's expressed concern with the privacy of chat data motivates a systematic investigation of privacy norms surrounding user chatbot data. Understanding such norms can help inform the design of chat interfaces as well as policy-level considerations for chat data.

The present work thus investigates chatbot privacy norms using the \textit{contextual integrity} framework, which posits that privacy norms exist in the context of \textit{information flows}. In short, an information exchange can be reduced to a set of transmission principles that govern the exchange of a certain kind of information (content) with a certain entity (role) for a certain reason (purpose)- collectively these influence whether the exchange of data is appropriate, or a cause for concern  \cite{nissenbaum2004privacy,malkin2022contextual}. Drawing on this theoretical framework, we set out to answer three research questions:

\begin{enumerate}
    \item \textbf{RQ1:} How do users perceive the sensitivity of chatbot data, and how does this compare to their perception of other forms of sensitive personal information? 
    \item \textbf{RQ2:} What factors (\textit{e.g.,} content, purpose, and consent, drawing on a contextual integrity framework) influence users' concerns about the handling of their chatbot data?
    \item \textbf{RQ3:} How willing are users to integrate chatbot data with data from their device, browser, and social accounts to improve their online experience - either via personalization or by gaining access to premium features?
\end{enumerate}

\noindent To address these questions, we drew on methods introduced in past work that addressed the contextual integrity of social media information flows \cite{gilbert2021measuring}, deploying a large-scale survey experiment with $N$=300 ChatGPT users residing in the U.S. In addition to a battery of questions that directly probe privacy attitudes, each respondent was also presented with 30 vignettes, short stories that each include a randomly drawn level from seven factors of interest, as illustrated in Figure \ref{fig:teaser}. These vignettes allowed us to isolate the factors that influence the two dependent variables rated by respondents: the \textit{appropriateness} of the information flow, and the level of \textit{concern} associated with it. Based on the results of our data analysis, we make three contributions:

\begin{enumerate}
    \item \textbf{Most respondents (82\%) consider chatbot conversations sensitive data, and on average agree that they have privacy concerns about ChatGPT conversations.} Respondents also disagreed on average that their conversations with chatbots like ChatGPT would remain private. Furthermore, respondents agreed on average that they were concerned about people receiving inaccurate information from AI; not understanding what AI can do; and having personal information misused by AI.
    \item \textbf{Only \textit{Informed Consent}, \textit{Anonymity}, and \textit{Privacy} (elements of a transmission principle) influence appropriateness perceptions.} Fitting linear mixed models (LMMs) to vignette ratings reveals that only providing informed consent, anonymizing data, and removing personal information improve respondents' perceptions of the appropriateness of the information flows described by our vignettes. The Role, Purpose, Content Type, and Location factors had no detectable effect on respondent perceptions - possibly because participants view data sharing as problematic regardless of recipient or purpose, or because the meaningful distinctions for users differ from those foregrounded by our experimental design.
    \item \textbf{Respondents uniformly \textit{reject} integrating personal data with chatbot data to improve their online experience.} Respondents on average disagreed that they would be willing to provide chatbots access to their search history, email, or device in exchange for either 1) better and more personalized chatbot features or 2) a premium-tier (\$200/month) chatbot subscription. Similarly, respondents prefer \textit{not} to have their chatbot data used to improve their social media experiences.
\end{enumerate}

\noindent Notably, despite their lack of confidence in the ultimate privacy of their chat data, many respondents reported discussing highly sensitive subjects with ChatGPT, including health and wellness (47.0\%), personal finances (35.3\%), and legal guidance (12.7\%), reflecting again the gap between expressions of concern and actual behavior observed in prior work \cite{zhang2024s,ngong2025protecting}.

Our findings have implications for AI policy and design. The primacy of a transmission protocol among our factors suggests the public would benefit from a systematic policy approach that establishes procedures around consent, anonymization, and handling of PII in chatbot data. Moreover, our third contribution suggests that developers may need to reconsider user willingness to exchange access to personal data for more advanced AI features, a key assumption for emerging agentic AI frameworks \cite{chan2023harms,chan2024visibility}.

\section{Related Work}

We first review the related work on the privacy-related attitudes and behaviors of users with respect to their chatbot data, and then introduce the contextual integrity framework. 

\subsection{Chatbot Privacy Attitudes and Concerns}

Building on a large body of research that seeks to understand the public's concerns about data privacy and emerging technologies like AI \cite{golda2024privacy}, recent work has applied both qualitative approaches and methods from computational social science to study users' attitudes toward the data they exchange with chatbots. \citet{alkamli2024understanding} collect and analyze Twitter data, finding that ChatGPT users express concern about 1) unauthorized access to their chat data and 2) their personal information being shared and exploited across platforms. \citet{ali2025understanding} analyze 2.5 million posts from the r/ChatGPT Reddit community, demonstrating ChatGPT users' concern about the collection and sharing of their personal data. Consistent with broader surveys of the public that demonstrate increases in concern about AI data privacy over time \cite{pew2023dataprivacy,pew2023aiprivacy}, \citet{ali2025understanding} found increases in the privacy awareness and sensitivity of ChatGPT users in response to reported security incidents and the rollout of new features with questionable effects on user privacy \cite{ali2025understanding}. Recent interview studies similarly find that users express significant concerns about their ability to permanently delete their chat data \cite{ma2025privacy}, and that they worry about the possibility of chat data being sold to data brokers or shared with third parties \cite{zhang2024s}. Concerns about the collection and secondary use of ostensibly private chat data have also motivated many organizations in domains ranging from medicine \cite{riedemann2024path} to journalism and fact-checking \cite{wolfe2024impact,wolfe2024implications} to adopt the use of \textit{open} chatbot models \cite{palmer2024using,paris2025opening,solaiman2023gradient}, rather than risking the security of client and patient data, or their own valuable proprietary information. 

We build on prior work by evaluating how chatbot users' stated privacy preferences vary across contexts. Establishing privacy norms that describe \textit{specific} data collected and exchanged in specific ways is essential for developing effective regulation for general-purpose AI applications like chatbots, which can be deployed across a wide variety of contexts and are known to ingest many forms of sensitive user data. \textbf{Our study thus examines the contextual factors that influence privacy expectations using an experimental survey method}, an approach that can isolate specific factors that influence user expectations for privacy.

\subsection{Chatbot User Privacy Behaviors}

Though studies of privacy attitudes find that users express uncertainty about the security of their chat data, research probing users' real-world \textit{behaviors} shows that users nonetheless disclose highly sensitive information to chatbots. Analysis of large, open-source collections of chatbot interactions (such as WildChat \cite{zhao2024wildchat}, LMSYS-Chat-1M \cite{zheng2023lmsys} or ShareGPT \cite{sharegpt52k}) find that users frequently discuss sensitive topics and share PII including email addresses, health conditions, passport numbers, and location data \cite{mireshghallah2024trust}. More targeted user studies report that even ``privacy conscious'' participants often indirectly disclose sensitive information \cite{ngong2025protecting}. Such disclosures have even prompted the development of privacy-preserving approaches to study the interactions that users engage in with chatbots, such as Anthropic's Claude insights and observations (Clio) platform \cite{tamkin2024clio}, which uses a layer of AI assistants to indirectly observe patterns in user data.

Prior work establishes two foundations on which we build: 1. Users' stated privacy preferences clash with their revealed preferences when interacting with chatbots; and, 2. Indirect disclosure of personal information is common across tasks, chatbot providers, and user intent. \textbf{Our study elicits users' stated preferences for privacy directly, as they rate the sensitivity of chatbot data against other data sources (\textit{e.g.}, email), and indirectly, by probing their willingness to integrate chatbot data across contexts.} 

\subsection{Contextual Integrity}
Contextual Integrity (CI) is both a theory and a diagnostic framework for understanding privacy in a networked information environment \cite{nissenbaum2019contextual}. As a theory, Nissenbaum and colleagues use CI to describe privacy as the emergent property of norms and practices that govern the appropriate exchange (or flow) of information \cite{barth2006privacy}. As a diagnostic framework, CI identifies five parameters in an information exchange: data type (the kind of information shared), data subject (what the information is about), sender (who is sharing the data), recipient (who receives the data), and transmission principle (the constraints or norms governing the flow of information) \cite{malkin2022contextual,kumar2024roadmap}. CI provides an ontology for making sense of the disruption of expectations when emerging technologies are adopted en masse \cite{vitak2023surveillance,apthorpe2018discovering}. Unsurprisingly, then, CI has been used in studies related to consumer facing AI \cite{mussgnug2024technology, cheng2024ci, ghalebikesabi2024operationalizing,fan2024goldcoin,li2024privacy}, including studies describing indirect disclosures in chatbot interactions \cite{ngong2025protecting}, and asymmetric privacy expectations between custom GPT creators and users \cite{ma2025privacy}. Building on prior work using factorial vignettes to understand CI norms \cite{martin2016measuring}, \citet{Mireshghallah2023CanLK} used CI-based vignettes to demonstrate that ChatGPT discloses private information in contexts where humans would not, highlighting a disconnect between the privacy reasoning of large language models and human privacy expectations.

We build on the use of CI in chatbot privacy research in two ways: 1. We deploy a factorial vignette survey based on CI's five parameters, focusing on exchanges of chatbot data; 2. We investigate the effectiveness of privacy interventions - anonymity, confidentiality, consent, and exchange (where a subject receives something in return) - as a transmission principle for chatbot data. \textbf{Our study yields evidence for a CI-based approach to chatbot privacy norms, with specific attention paid to CI's transmission principle.}
\begin{table*}[t]
\renewcommand{\arraystretch}{1.2}
\centering
\begin{tabular}{p{4cm}p{12cm}}
\hline
\multicolumn{2}{p{16cm}}{
    \textbf{Vignette Format:} Your ChatGPT conversations were shared with a(n) \textless{}Role\textgreater{} based in \textless{}Location\textgreater{} for the purpose of \textless{}Purpose\textgreater{}. \textless{}Content\textgreater{} were shared. You were \textless{}Consent/Awareness\textgreater{}. Your conversations were \textless{}Anonymity\textgreater{} before being shared. Personally identifying information from your conversations was \textless{}Privacy\textgreater{}.
} \\ 
\midrule
\textbf{Factor} & \multicolumn{1}{c}{\textbf{Levels}} \\
\midrule
\colorbox{green!40}{Role} & A big tech company; A hospital; A government agency; An insurance company; A university computer science researcher; A university social science researcher; A charitable foundation \\
\colorbox{green!40}{Location} & the United States; the European Union; China \\
\colorbox{green!40}{Purpose} & training future AI models; creating a public dataset for AI research; improving user experience with AI; fighting terrorism; assessing mental health; personalizing advertising; predicting human behavior; studying the risks of AI to society \\
\colorbox{cyan!40}{Content} & all of your conversations with ChatGPT; those conversations with ChatGPT where you used the model to help with your job; those conversations with ChatGPT about your social life and personal relationships; those conversations with ChatGPT about your personal health and wellness; those conversations with ChatGPT where you sought legal or ethical guidance \\
\colorbox{yellow!40}{Consent/Awareness} & asked for consent in advance of the data being collected; informed that your data was collected; not be informed that your data was collected \\
\colorbox{yellow!40}{Anonymity} & anonymized; not anonymized \\
\colorbox{yellow!40}{Privacy} & automatically removed; not removed \\ \hline
\end{tabular}
\caption{The format and factors of the factorial vignettes presented in part 3 of our survey instrument. Each vignette is constructed by randomly selecting a level from each of seven factors, and each participant is presented with a total of 30 vignettes.}
\label{tab:vignette-factors}
\end{table*}

\section{Approach}

We describe the development and deployment of the survey instrument used to address our research questions. We then characterize our participants and our approach to analyzing the survey data we collected. The Institutional Review Board (IRB) at our University reviewed and approved this research.

\subsection{Survey Instrument and Data Analysis}

We used the Qualtrics survey design software to create a five-part survey, described in detail below.

\noindent \textbf{Part 1: Privacy Attitudes.} We first asked participants five questions about their attitudes towards general information privacy topics (\textit{e.g.,} ``I am concerned that online companies are collecting too much information about me'') and about ChatGPT in particular (\textit{e.g.,} ``I have privacy concerns about my conversations with ChatGPT''). We also asked respondents to rate their level of concern about the public 1) getting inaccurate information from AI; 2) not understanding what AI can do; and 3) having personal information misused by AI. For general privacy concerns we adopt questions asked in \citet{gilbert2021measuring}, and for AI concerns among the public we draw questions from \citet{pew2023aiprivacy}, allowing comparison between studies. Participants responded on a 100-point scale where 0 indicated strong disagreement, 50 neutral, and 100 strong agreement.

\noindent \textbf{\textit{Data Analysis.}} For each question, we report the mean and standard deviation and use a one-sample $t$-test of significance against a hypothesized mean of 50 (neutral).

\noindent \textbf{Part 2: Private Data Exchange Value.} Next, we asked participants about their willingness to exchange access to their search history, emails, device data, or chatbot conversations for 1) personalized chatbot responses and insights; and 2) a premium chatbot subscription valued at \$200 (analogous to ChatGPT Pro \cite{openai2024chatgptpro}). We then asked participants if they would provide access to their chatbot history to improve services like search results, product recommendations, social media post popularity, and social media feed moderation. We used the same 100-point scale as Part 1.

\noindent \textbf{\textit{Data Analysis.}} For each question, we report the mean and standard deviation, and we use a one-sample $t$-test against a hypothesized mean of 50 (neutral).

\begin{table*}
\small
\setlength{\tabcolsep}{6pt}
\renewcommand{\arraystretch}{1.2}
\centering
\begin{tabularx}{.99\textwidth}{p{2cm}p{1cm}p{2cm}p{1cm}p{.75cm}p{.5cm}p{2cm}p{1cm}p{2cm}p{1cm}}
\toprule
\textbf{Gender} & & \textbf{Education} & & \textbf{Age} & & \textbf{Politics} & & \textbf{Race} & \\
\midrule
Man & 54.0\% & HS or less & 10.7\% & Avg & 36 & Democrat & 48.7\% & White & 62.3\%  \\ 
Woman & 43.3\% & Associate's & 11.0\% & SD & 10.8 & Republican & 23.0\% & Black  & 16.0\% \\
Non-Binary & 0.7\% & Some college & 22.7\% & Min & 18 & Independent & 20.3\% & Asian & 9.0\% \\
Something Else & 1.3\% & Bachelor's & 42.7\% & Max & 76 & Something else & 5.7\% & Pacific Islander & 0.6\% \\
Prefer not say & 0.7\% & Master's & 9.3\% & & & Prefer not say & 2.3\% & Indigenous & 2.0\% \\
& & Professional & 1.7\% & & & & & Something Else & 3.7\% \\
& & Doctorate & 1.7\% & & & & & Other & 3.3\% \\
& & Prefer not say & 0.3\% & & & & & Prefer not say & 1.0\% \\
\bottomrule
\end{tabularx}
\caption{Participant demographics for the $N$=300 respondents who completed our full survey instrument. Our population over-represented Men, Democrats, Middle-Aged individuals, and College-Educated individuals relative to the U.S. nationally. }
\label{tab:participants}
\end{table*}

\begin{table*}
\small
\setlength{\tabcolsep}{4pt}
\renewcommand{\arraystretch}{1.2}
\centering
\begin{tabularx}{\textwidth}{@{}lX lX lX lX@{}}
\toprule
\textbf{Year Started Using} & & \textbf{Usage Frequency} & & \textbf{Account Type} & \\
\midrule
2023 or earlier & 36.9\% & Daily & 28.6\% & No Account & 16.2\% & \\
2024 & 57.6\% & Weekly & 46.9\% & ChatGPT Free Subscription & 72.4\% & \\
2025 & 9.0\%  & Less than Monthly & 15.5\% & ChatGPT Plus & 12.8\% \\
& & Monthly & 12.4\% &  ChatGPT Pro & 2.1\% \\
\bottomrule
\end{tabularx}
\caption{ChatGPT usage among our respondents, who mostly use Free-tier accounts and mostly use ChatGPT at least weekly.}
\label{tab:participant-chatgpt-usage}
\end{table*}

\noindent \textbf{Part 3: Factorial Vignettes.} Next, participants were shown paragraph-length scenarios (vignettes) describing the use of their ChatGPT conversation history. Each vignette followed the format described in the header of Table \ref{tab:vignette-factors}, with the seven factors denoted in $<>$ tags each randomly replaced with a level included in the table. We drew the Role, Purpose, Content, and Consent/Awareness factors from prior work using factorial vignettes to describe the contextual integrity of information flows \cite{martin2012diminished, gilbert2021measuring}. We added the Location factor to account for the newly salient geopolitical dimension of AI data privacy \cite{deepseekshock2025,act2024eu}, and we added the Anonymity and Privacy factors because much recent work notes that many users exchange sensitive and personally identifying data during chatbot conversations \cite{belen2021privacy}. In total, each participant judged 30 vignettes on two dimensions: ``This is an \textbf{appropriate} use of my ChatGPT data'' (dependent variable 1); and 2. ``This use of my data would \textbf{concern} me'' (dependent variable 2). In total, the $N$=300 participants in our survey each judged 30 vignettes, yielding 9000 judgments about the use of chatbot data on two dimensions - Appropriateness and Concern.

\noindent \textbf{\textit{Data Analysis.}} We analyze vignette responses by fitting two linear mixed models (LMMs) with Appropriateness (dependent variable 1) and Concern (dependent variable 2). Each LMM is fit to the seven factors described in Table \ref{tab:vignette-factors}, along with a participant random effect. We set the baseline level for the factor \textit{Role} to be ``a hospital,'' for \textit{Purpose} to be ``improving user experience with AI,'' for \textit{Location} to be ``the US,'' for \textit{Content} to be ``all conversations with ChatGPT'' and for \textit{Consent} to be ``asked for consent in advance of data collection.'' Baseline levels were automatically selected for remaining factors by the LMER package in R, as these factors were either binary or had no clear baseline. Because our models failed assumptions for the homoscedasticity of residuals based on Kolmogorov-Smirnov tests, we used a cluster-robust covariance matrix with CR1 estimate to adjust for our data's multilevel structure \cite{cameron2010robust}. 

In addition to the base LMMs, we also compute interaction effects between the five privacy attitude questions (Survey part 1) with 1) the Consent factor and 2) the Location factor. We again fit separate LMMs to the Appropriateness and Concern dependent variables, examining only interactions with Consent and Location, rather than all factors.

\noindent \textbf{Part 4: ChatGPT Usage.} Next, we ask participants about their ChatGPT usage, including their account type, date of first use, frequency of use, tasks they typically complete, and topics they frequently discuss. For the last two questions, respondents could select multiple options.

\noindent \textbf{\textit{Data Analysis.}} We report summary statistics as a percentage of participant responses for each question. 

\noindent \textbf{Part 5: Chat Data Sensitivity.} Finally, we ask respondents to rank the sensitivity of 14 discrete forms of personal information, including their social security number, phone calls, emails, and conversations with chatbots like ChatGPT. Respondents were asked to classify each form of information as ``Highly Sensitive,'' ``Sensitive,'' or ``Not Sensitive.''

\noindent \textbf{\textit{Data Analysis.}} We use a Friedman Test to check for differences in the sensitivity of the forms of personal information. We then use Wilcoxon signed-rank tests with Bonferroni correction to test for differences between chatbot conversation sensitivity and sensitivity of other forms of information.

\subsection{Participants}

We recruited survey respondents via CloudResearch Connect \cite{hartman2023introducing}. Our inclusion criteria required participants to use ChatGPT monthly, be 18 years or older, and reside in the United States. Table \ref{tab:participants} describes the demographic composition of our sample, while Table \ref{tab:participant-chatgpt-usage} describes ChatGPT use among our sample. Most respondents were frequent ChatGPT users, with 46.9\% using the application at least weekly and 28.6\% using it at least daily.

Participants were required to answer all questions in the survey. The mean time to complete the survey was 21 minutes, 33 seconds (SD of 12 minutes, 45 seconds), and the median was 17 minutes, 30 seconds. We manually inspected survey responses completed in less than 10 minutes and discarded those completed in less than 9 minutes.

\subsection{Preregistration}

We submitted a pre-registration of our hypotheses and planned data analysis with the Open Science Foundation in April 2025. This paper is consistent with that preregistration, but we made several significant changes to our modeling and hypothesis testing that prompted us to submit a Statement of Transparent Changes, following best practices in research reporting \cite{nosek2018preregistration, lakens2024and}. The original pre-registration and the Statement of Transparent Changes can be view here: \url{https://osf.io/m6jt5/}
\section{Results}


\begin{table}[h]
\centering
\small
\begin{tabular}{p{3.6cm}cccc}
\toprule
\multicolumn{5}{c}{\textbf{Privacy Attitudes}} \\
\midrule
\textbf{Statement} & \textbf{Mean} & \textbf{SD} & \textbf{t} & \textbf{p $<$} \\
\midrule
Online companies collect too much personal info & 74.9 & 22.4 & 19.3 & 0.01 \\
In general, I trust websites & 43.8 & 24.6 & -4.4 & 0.01 \\
In general, I believe privacy is important & 88.7 & 15.1 & 44.4 & 0.01 \\
Privacy concerns about conversations with ChatGPT & 63.6 & 27.9 & 8.5 & 0.01 \\
My chats with bots like ChatGPT will stay private & 40.0 & 27.6 & -6.9 & 0.01 \\
\bottomrule
\end{tabular}
\caption{Participants disagreed with statements that they trust websites and believe their chat data will stay private, while agreeing with statements that they have privacy concerns about ChatGPT data and that companies collect too much personal information.}
\label{tab:privacy-attitudes}
\end{table}

\subsection{Privacy Attitudes}

As seen in Table \ref{tab:privacy-attitudes}, our results unambiguously show that participants feel concern about the privacy of their chat data and online data more broadly, as they agreed with questions stating 1) that online companies are collecting too much personal information and 2) that they have privacy concerns about their conversations with ChatGPT. Conversely, participants disagreed with questions stating 1) that they trust websites, and 2) that they believe their conversations with chatbots like ChatGPT will remain private. Based on \citet{gilbert2021measuring}, we hypothesized that we would observe statistically significant agreement with questions like online companies collect too much personal information'' and in general, I believe privacy is important'', with means between 50 and 60. However, we observe notably higher means and tt
t-values than expected.

In Table \ref{tab:ai-concerns}, we observe agreement with all three questions probing participants' concerns about AI specifically, reflecting concerns about receiving inaccurate information from AI, not understanding what AI can do, and personal data being misused by AI. These responses are high in magnitude and consistent across political demographic groups (\textit{e.g.,} Republicans, Democrats), consistent with large-scale surveys about the extent to which AI concerns are shared across society \cite{pew2023aiprivacy}. Taken together, our participants' responses add to studies indicating that Americans' concern about data privacy has grown significantly over recent years \cite{pew2023dataprivacy}, including with respect to data exchanged with AI.

\begin{table}[h]
\centering
\small
\begin{tabular}{p{3.5cm}cccc}
\toprule
\multicolumn{5}{c}{\textbf{AI Concerns Question Battery}} \\
\midrule
\textbf{Concern} & \textbf{Mean} & \textbf{SD} & \textbf{t} & \textbf{p $<$}\\
\midrule
People getting inaccurate information from AI & 72.8 & 21.8 & 18.1 & 0.01\\
People not understanding what AI can do & 68.2 & 24.5 & 12.9 & 0.01\\
People's personal info being misused by AI & 72.7 & 24.3 & 16.2 & 0.01\\
\bottomrule
\end{tabular}
\caption{Responses to ``When it comes to artificial intelligence, how concerned are you about...[\textbf{Concern}]?'' demonstrates consistent concerns about AI data privacy and quality.}
\label{tab:ai-concerns}
\end{table}

\begin{table}[h]
\centering
\small
\begin{tabular}{p{3.5cm}cccc}
\toprule
\multicolumn{5}{c}{\textbf{Willingness to Share Personal Data for Personalization}} \\
\midrule
\textbf{Information Type} & \textbf{Mean} & \textbf{SD}& \textbf{t} & \textbf{p$<$} \\
\midrule
Chatbot History & 62.9 & 27.4 & 8.0 & 0.01 \\
Search History & 34.9 & 29.6 & -8.4 & 0.01 \\
Email & 16.6 & 23.8 & -18.7 & 0.01 \\
Device & 21.4 & 26.3 & -16.5 & 0.01\\
\bottomrule
\end{tabular}
\caption{Responses to ``I would be willing to give chatbots access to my \textbf{[Information Type]} in exchange for more personalized responses and insights'' show that most respondents would not share personal data to improve a chatbot.}
\label{tab:personalization-prefs}
\end{table}

\begin{table}[h]
\centering
\small
\begin{tabular}{p{3.5cm}cccc}
\toprule
\multicolumn{5}{c}{\textbf{Willingness to Share Personal Data for Premium Chatbot}} \\
\midrule
\textbf{Information Type} & \textbf{Mean} & \textbf{SD} & \textbf{t} & \textbf{p $<$}\\
\midrule
Chatbot History & 61.1 & 33.6 & 5.7 & 0.01\\
Search History & 36.8 & 33.7 & -6.8 & 0.01\\
Email & 24.7 & 32.1 & -13.7 & 0.01\\
Device & 26.7 & 32.4 & -12.4 & 0.01\\
\bottomrule
\end{tabular}
\caption{Responses to ``In exchange for a premium chatbot subscription — which is valued at \$200 a month and includes priority access to new features — I would be willing to give access to my...[\textbf{Information Type}]?'' show that even access to premium features does not induce respondents to share their personal data with a chatbot.}
\label{tab:premium-chatbot}
\end{table}

\subsection{Private Data Exchange}

As described in Table \ref{tab:personalization-prefs}, the only form of data participants were willing to share with chatbots in order to receive more personalized responses and insights was their chatbot history; participants were not willing to share access to their search history, email, or on-device data and applications in exchange for these features. These results were consistent when participants were instead presented with the option to exchange their data for a premium (\$200/month) chatbot subscription, as described in Table \ref{tab:premium-chatbot}. This is noteworthy given the incorporation of LLM-based chatbots into operating systems (as in the case of assistants like Microsoft's Copilot), mobile devices (as with ChatGPT for iPhone), and web interfaces, most notably AI agents capable of carrying out actions on behalf of the user. While the public could conceivably be willing to provide access to data for emerging LLM-based technologies in exchange for notable new capabilities, our results suggest that Americans are \textit{not} willing to make that trade simply to equip a chatbot assistant with personalized features, even if those features are valued at \$200. 

\begin{table}[h]
\centering
\small
\begin{tabular}{p{3.5cm}cccc}
\toprule
\multicolumn{5}{c}{\textbf{Cross-Application Integration of Chat History}} \\
\midrule
\textbf{Use of Chat History} & \textbf{Mean} & \textbf{SD} & \textbf{t} & \textbf{p $<$}\\
\midrule
Search engine results & 47.7 & 32.0 & -1.2 & 0.01\\
Product recommendations & 46.9 & 32.9 & -1.6 & 0.01\\
Popularity of my posts on social media & 26.6 & 30.3 & -13.4 & 0.01\\
Content moderation on my social media & 32.6 & 30.7 & -9.8 & 0.01\\
\bottomrule
\end{tabular}
\caption{Responses to ``I would be willing to have my conversation history used to improve...[\textbf{Use of Chat History}]'' demonstrate little enthusiasm for integrating chat data to improve other applications, particularly on social media.}
\label{tab:secondary-use}
\end{table}

As shown in Table \ref{tab:secondary-use}, respondents were ambivalent about their chatbot history being used to improve search results or product recommendations, as reflected in means just below 50 and large standard deviations. However, respondents were \textit{not} willing to use their chatbot history to improve their social media experience. Providers like Meta and X have implemented LLM-based chatbots on their platforms \cite{xai2025grok3,metaai}, but our results suggest that respondents prefer to preserve the independence of their chat data from their social networks, even if it is useful for their experience

\begin{table*}[t]
\centering
\small
\begin{tabular}{p{7.5cm}|rrrr|rrrr}
\toprule
\multicolumn{1}{c|}{\textbf{Factors}} & \multicolumn{4}{c|}{\textbf{Appropriateness LMM}} & \multicolumn{4}{c}{\textbf{Concern LMM}} \\
\midrule
\midrule & Est & \textit{SE} & \textit{t} & \textit{p} & Est & \textit{SE} & \textit{t} & \textit{p} \\
\textbf{Intercept} \ & 41.38 & 1.69  & 24.48 & $<$0.01 & 62.36 & 1.76 & 35.40 & $<$0.01 \\
\midrule
\midrule
\textbf{Role} \ (Intercept: Hospital) & Est & \textit{SE} & \textit{t} & \textit{p} & Est & \textit{SE} & \textit{t} & \textit{p} \\
\midrule
Big tech company & 1.28 & 1.07 & 1.20 & 0.23 & 0.18 & 1.12 & 0.16 & 0.87 \\
University computer scientist & 1.04 & 1.15 & 0.90 & 0.37 & 0.92 & 1.08 & 0.85 & 0.39 \\
University social science researcher & 0.31 & 1.05 & 0.29 & 0.77 & 1.54 & 1.15 & 1.34 & 0.18 \\
Charitable Foundation & 0.12 & 1.16 & 0.11 & 0.92 & 1.61 & 1.21 & 1.32 & 0.19 \\
Government agency & -0.23 & 1.09 & -0.21 & 0.83 & 0.69 & 1.16 & 0.59 & 0.55 \\
Insurance company &0.32 & 1.12 & 0.29 & 0.77 & 1.26 & 1.10 & 1.15 & 0.25 \\
\midrule
\textbf{Purpose} \ (Intercept: Improving user experience with AI) & Est & \textit{SE} & \textit{t} & \textit{p} & Est & \textit{SE} & \textit{t} & \textit{p} \\
\midrule
Assessing mental health & -0.95 & 1.07 & -0.87 & 0.38 & 1.02 & 1.07 & 0.95 & 0.34 \\
Creating a public dataset for AI research & 0.13 & 1.05 & 0.12 & 0.90 & 0.39 & 1.09 & 0.35 & 0.72 \\
Fighting Terrorism & -1.41 & 1.11 & -1.27 & 0.21 & 0.50 & 1.12 & 0.45 & 0.66 \\
Personalizing Advertising & -0.71 & 1.17 & -0.61 & 0.54 & 0.08 & 1.17 & 0.07 & 0.94 \\
Predicting human behavior & -0.47 & 1.13 & -0.42 & 0.68 & 1.04 & 1.16 & 0.89 & 0.37 \\
Studying AI risks to society & 0.41 & 1.09 & 0.37 & 0.71 & 0.02 & 1.08 & 0.02 & 0.98 \\
Training future AI models & -0.38 & 1.16 & -0.33 & 0.74 & 0.54 & 1.13 & 0.48 & 0.63 \\
\midrule
\textbf{Content Type} \ (Intercept: All conversations with ChatGPT) & Est & \textit{SE} & \textit{t} & \textit{p} & Est & \textit{SE} & \textit{t} & \textit{p} \\
\midrule
Conversations using ChatGPT to help with your job & 0.03 & 0.94 & 0.03 & 0.97 & 0.88 & 0.89 & 0.98 & 0.33 \\
Conversations about social life and personal relationships & -1.50 & 0.82 & -1.82 & 0.07 & 1.31 & 0.88 & 1.49 & 0.14 \\
Conversations about your personal health and wellness & -0.44 & 0.86 & -0.51 & 0.61 & 1.26 & 0.89 & 1.42 & 0.16 \\
Conversations where you sought legal or ethical guidance & 1.47 & 1.01 & 1.46 & 0.14 & 0.17 & 1.00 & 0.17 & 0.87 \\
\midrule
\textbf{Location} \ (Intercept: United States) & Est & \textit{SE} & \textit{t} & \textit{p} & Est & \textit{SE} & \textit{t} & \textit{p} \\
\midrule
China & 1.28 & 1.07 & 1.20 & 0.23 & 0.81 & 0.70 & 1.15 & 0.25 \\
The European Union & 1.04 & 1.15 & 0.90 & 0.37 & -0.02 & 0.73 & -0.02 & 0.98 \\
\midrule
\textbf{Consent} \ (Intercept: Asked for consent in advance) & Est & \textit{SE} & \textit{t} & \textit{p} & Est & \textit{SE} & \textit{t} & \textit{p} \\
\midrule
\cellcolor{gray!30} Not informed that your data was collected & \cellcolor{gray!30} -4.15 & \cellcolor{gray!30} 0.76 & \cellcolor{gray!30} -5.46 & \cellcolor{gray!30} $<$\textbf{.001} & \cellcolor{gray!30} 2.42 & \cellcolor{gray!30} 0.73 & \cellcolor{gray!30} 3.34 & \cellcolor{gray!30} $<$\textbf{.001}\\
Informed that your data was collected & -0.37 & 0.71 & -5.42 & 0.60 & 0.23 &  0.71 & 0.32 & 0.75 \\
\midrule
\textbf{Anonymity} \ (Intercept: Anonymized) & Est & \textit{SE} & \textit{t} & \textit{p} & Est & \textit{SE} & \textit{t} & \textit{p} \\
\midrule
\cellcolor{gray!30} Not anonymized & \cellcolor{gray!30}-1.36 & \cellcolor{gray!30}0.58 & \cellcolor{gray!30}-2.36 & \cellcolor{gray!30} $<$\textbf{.001} & 1.12 & 0.58 & 1.92 & 0.06 \\
\midrule
\textbf{Privacy} \ (Intercept: PII Removed) & Est & \textit{SE} & \textit{t} & \textit{p} & Est & \textit{SE} & \textit{t} & \textit{p} \\
\midrule
\cellcolor{gray!30} PII not removed & \cellcolor{gray!30} -2.03 & \cellcolor{gray!30}0.59 & \cellcolor{gray!30} -3.48 & \cellcolor{gray!30} $<$\textbf{.001} & \cellcolor{gray!30} 2.33 & \cellcolor{gray!30} 0.58 & \cellcolor{gray!30} 3.98 & \cellcolor{gray!30} $<$\textbf{.001} \\

\bottomrule
\end{tabular}
\caption{Significant LMM results for Consent, Anonymity, and Privacy with Appropriateness, and Consent and Privacy with Concern, indicate that variation in perceptions of chatbot information flows depends primarily on the Transmission Principle.}
\label{tab:full-model}
\end{table*}

\subsection{Factorial Vignettes}

To analyze the 9000 factorial vignette judgments made by our survey participants, we fit two Linear Mixed Models to the 1) Appropriateness and 2) Level of Concern dependent variables. As shown in Table \ref{tab:full-model}, our findings indicate that only the Consent and Anonymity factors have a significant effect on the Appropriateness variable, while the Consent and Privacy factors have a significant effect on the Level of Concern variable. Though Anonymity has a large $t$-value in the Concern model, its $p$-value narrowly misses the threshold for significance ($p$=.06). While results for the Privacy, Anonymity, and Consent factors align with our hypotheses, we found that our expectations about the remaining factors were not borne out. We expected that our American survey respondents would report lower levels of concern and higher perceived appropriateness for chat data exchanged with U.S. entities (and vice-versa for European Union and China), but our model indicates that there is no significant difference between the levels of the Location factor for either dependent variable. Similarly, though we expected to observe higher levels of concern when chat data was exchanged with entities such as an insurance company or a big tech company, the model indicates no statistically discernible difference between the levels of the Role factor. In keeping with our Privacy Attitudes results, the grand means of our dependent variables across all vignettes reflect a sense of concern about the exchange of chat data, as the mean for Appropriateness was 37.6 ($\sigma$=31.3), and for Concern was 67.3 ($\sigma$=30.7).

\begin{figure*}
\centering
\begin{tikzpicture}
\begin{axis}[
    xbar stacked,
    bar width=8.1pt,
    xmin=-2.25,
    xmax=102.25,
    xtick=\empty,
    ytick=\empty,
    width=14cm,
    height=9cm,
    title={\large Perceived Sensitivity of Personal Information},
    xlabel={\% of Responses},
    ticklabel style={font=\small},
    symbolic y coords={Political Views, Media Preferences, Content of Emails, Content of Social Media Posts, Religious Views, Purchasing Habits, Chatbot Conversations, Websites Visited, Content of Phone Conversations, Who Friends Are and What They Like, Location Over Time, Content of Text Phone Conversations, State of Health, Social Security Number},
    ytick=data,
    legend style={at={(0.86,0.23)}, anchor=north, legend columns=1, font=\small},
    nodes near coords,
    nodes near coords align={horizontal},
    nodes near coords style={text=white,xshift=-6pt}
]
\addplot+[xbar, fill=red!70] coordinates {(3,Political Views) (6,Media Preferences) (7,Content of Emails) (8,Content of Social Media Posts) (12,Religious Views) (18,Purchasing Habits) (22,Chatbot Conversations) (22,Websites Visited) (57,Content of Phone Conversations) (57,Who Friends Are and What They Like) (65,Location Over Time) (68,Content of Text Phone Conversations) (68,State of Health) (98,Social Security Number)};
\addplot+[xbar, fill=orange!60] coordinates {(27,Political Views) (32,Media Preferences) (34,Content of Emails) (39,Content of Social Media Posts) (43,Religious Views) (49,Purchasing Habits) (60,Chatbot Conversations) (62,Websites Visited) (41,Content of Phone Conversations) (39,Who Friends Are and What They Like) (33,Location Over Time) (30,Content of Text Phone Conversations) (28,State of Health) (2,Social Security Number)};
\addplot+[xbar, fill=gray!60] coordinates {(70,Political Views) (62,Media Preferences) (59,Content of Emails) (53,Content of Social Media Posts) (45,Religious Views) (33,Purchasing Habits) (18,Chatbot Conversations) (16,Websites Visited) (2,Content of Phone Conversations) (4,Who Friends Are and What They Like) (2,Location Over Time) (2,Content of Text Phone Conversations) (4,State of Health) (0,Social Security Number)};
\legend{Highly Sensitive, Sensitive, Not Sensitive}
\end{axis}
\end{tikzpicture}
\caption{82\% of respondents viewed chatbot data as sensitive or highly sensitive. Chat data was perceived as more sensitive than email contents, social media posts, and religious or political views, but less sensitive than text or verbal phone conversations.}
\label{fig:latex-sensitivity}
\end{figure*}
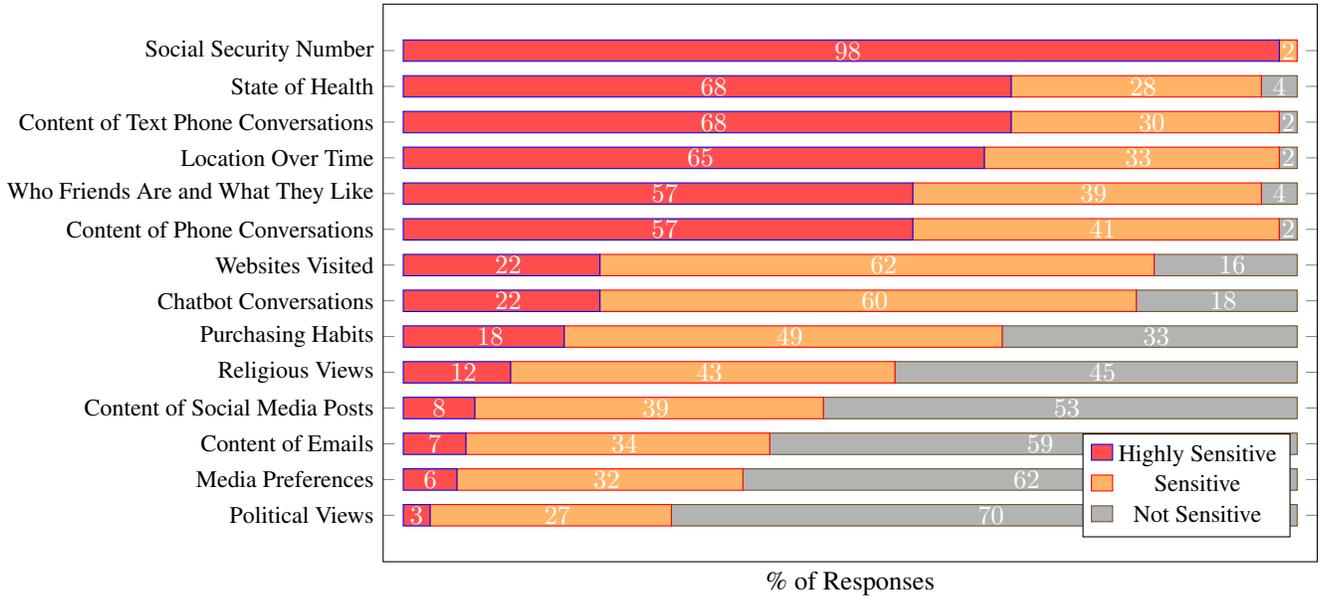

\noindent \textbf{Interaction Effects.} In accordance with our hypotheses, we tested interactions for responses to our Privacy Attitudes questions with 1) the Consent factor and 2) the Location factor. Note that while we fit the interactions to the full model (with all factors), we only examined results for Location and Consent in accordance with our hypotheses, and to reduce the risk of Type 1 errors. Though we hypothesized that responses to the Privacy Attitudes questions indicating greater privacy concern would interact with the Consent/Awareness factor, such that we would observe increases in the level of concern when informed consent was not obtained, we did not observe any such interaction. As clearly indicated from our main effects, Consent/Awareness exerts a significant influence on perceptions of both appropriateness and concern, but the interaction model results indicate that this is not determined by participants' self-reported privacy attitudes. We include full results for interaction effects in the Appendix.

\begin{table}[h]
\centering
\small
\begin{tabular}{p{4.6cm}cccc}
\toprule
\multicolumn{3}{c}{\textbf{Topics Discussed with ChatGPT}} \\
\midrule
\textbf{Topic} & \textbf{Count} & \textbf{Percentage}\\
\midrule
Work/Job Support & 159 & 53 \% \\
Scientific or Academic Information & 159 & 53 \% \\
Health or Wellness Advice  & 141 & 47 \% \\
Entertainment & 119 & 40 \% \\
Current Events & 118 & 40 \% \\
Hobbies & 106 & 35 \% \\
Travel & 106 & 35 \% \\
Personal Finances & 106 & 35 \% \\
Mental Health Support & 80 & 26 \% \\
Personal Relationships & 62 & 21 \% \\
Politics & 57 & 19 \% \\
Legal Guidance & 38 & 13 \% \\
Religion & 28 & 9 \% \\
Sexual \& Erotic & 17 & 6 \% \\
Other & 20 & 7 \% \\
\bottomrule
\end{tabular}
\caption{Substantial pluralities of participants discussed sensitive topics such as health and wellness, personal finances, mental health, and legal guidance with ChatGPT, despite their reservations about the privacy of this data.}
\label{tab:chatgpt-topics}
\end{table}

\subsection{ChatGPT Usage}

Table \ref{tab:chatgpt-topics} describes the most common topics discussed with ChatGPT by our respondents. More than half use ChatGPT to discuss academic or scientific topics, and to discuss topics relevant to supporting their job. More important for this study, 47.0\% of respondents used ChatGPT for Health and Wellness Advice, while 35.3\% used the model for Personal Finances, and 26.7\% used it for Mental Health Support, and 12.7\% for Legal Guidance. While information about any of the topics we asked about could conceivably be used to make valuable inferences about a user, discussions of these topics present clear vectors for the exchange of sensitive personal information. It is thus notable that so many respondents willingly discuss such sensitive topics with ChatGPT, despite the clear reservations about chatbot privacy reported in our survey's previous question batteries (\textit{e.g.}, Privacy Attitudes).

Table \ref{tab:chatgpt-tasks} reports the most common tasks for which ChatGPT was used by our respondents. Searching for Information was the most common task, selected by very nearly the entirety of our sample, and perhaps indicating why, in our subsequent analysis, the perceived sensitivity of chat conversations is analogous to that of one's browsing history. We also note that, compared to other studies of user chatbot data, our participants self-report substantially lower uses of chatbots for creative work, sexual roleplay, and software development, which may speak to a gap between stated and revealed preferences \cite{tamkin2024clio, longpre2024consent}. 

\subsection{Chat Data Sensitivity}
Our  respondents' perceived sensitivity of various forms of personal information are described in Figure \ref{fig:latex-sensitivity}. A Friedman test revealed a statistically significant difference in perceived sensitivity between information types, with $\chi^2(13) = 2160.5$, $p < 0.01$. Post-hoc tests with Bonferroni corrections further indicated significant differences between the perceived sensitivity of chatbot conversations and each of the other forms of personal information, with the notable exception of records of websites visited online (\textit{i.e.,} browsing history), for which the post-hoc comparison was not significant. Respondents were most likely to rank their Social Security number, details of their physical location over a period of time, the state of their health, and the content of their texts and phone conversations as sensitive or highly sensitive. Though chatbot conversations were less likely to be rated as highly sensitive than these forms of information, more than 80\% of respondents rated chatbot conversations as sensitive or highly sensitive, with the majority (60\%) characterizing this data as sensitive. Chatbot conversations were perceived as more sensitive than the contents of email messages or social media posts, but less sensitive than text and phone conversations, suggesting chatbot interactions may occupy an intermediate space between information exchanges that are more formal (email) or performative (social media) and those that are more private and personal (text and phone).

\begin{table}[h]
\centering
\small
\begin{tabular}{p{4.6cm}cccc}
\toprule
\multicolumn{3}{c}{\textbf{Tasks ChatGPT is Used For}} \\
\midrule
\textbf{Task} & \textbf{Count} & \textbf{Percentage}\\
\midrule
Searching for Information & 288 & 96 \% \\
Writing and Editing & 196 & 65 \% \\
Asking for Advice  & 155 & 52 \% \\
Personal Education (learning) & 143 & 48 \% \\
Daily Life Skills (cooking, finances) & 89 & 30 \% \\
Computer Programming & 87 & 30 \% \\
Art and Design & 49 & 16 \% \\
Other & 20 & 7 \% \\
\bottomrule
\end{tabular}
\caption{96\% of respondents used ChatGPT to search for information, suggesting a reason that respondents also rate chat data as having similar sensitivity to browsing history.}
\label{tab:chatgpt-tasks}
\end{table}
\begin{table*}[t]
\centering
\small
\begin{tabular}{p{5.5cm}|p{5.5cm}|p{5.5cm}}
\toprule
\multicolumn{1}{c|}{\textbf{Research Question}} & \multicolumn{1}{c|}{\textbf{Finding}} & \multicolumn{1}{c|}{\textbf{Implication}} \\
\midrule \midrule
\textbf{RQ1:} How do users perceive the sensitivity of chatbot data relative to other forms of personal information? & 80+\% of respondents view chat data as sensitive (less sensitive than a phone call but more sensitive than an email).  &  Chat data is  disclosive, and a gap in reported sensitivity may indicate users' inability to recognize privacy tradeoffs. \\
\midrule

\textbf{RQ2:} What factors influence users' concern with handling of their chat data?  & Consent, anonymization, and removal of PII (\textit{i.e.,}  features of a transmission principle in contextual integrity) are significant in our model. & For an emerging technology like LLM-based chatbots, users judge information exchanges based on governing principles that are well understood and often used. \\

\midrule
\textbf{RQ3:} Are users willing to exchange their personal data for premium or personalized chatbot features? & No, respondents reject exchanging data (emails, search histories, device access) for premium and personalized AI features. & AI companies should not assume willingness to integrate data across contexts for more advanced models (\textit{e.g.}, agentic AI). \\

\bottomrule
\end{tabular}
\caption{A Summary of Our Primary Research Questions, Findings, and Implications.}
\label{tab:implications}
\end{table*}

\section{Discussion}
Our research reflects both growing concern about data privacy online and declining trust for data protections \cite{pew2023aiprivacy, pew2023dataprivacy}. The societal consequences are identifiable in our findings. Consider that \textit{none} of the levels for our Role factor exhibited significant effects on appropriateness or concern, such that we observe no benevolence paid to ostensibly responsible actors such as university researchers or charitable foundations. Similarly, we observed no significant effects for the Purpose factor, such that respondents were not swayed when scenarios included studying AI risks, fighting terrorism, or creating public datasets. We also found no support for our hypothesis that American respondents would be more concerned about foreign (China and European) receivers of their chat data than domestic ones. Our results suggest participants view chatbot data sharing as problematic, regardless of content, recipient, or stated purpose, particularly when exchanges occur without informed consent, anonymization, and proper removal of sensitive data. We acknowledge that meaningful distinctions for users may exist that differ from those captured in our experimental design, which focuses on hypothetical vignettes. But on average, our participants \textit{disagreed} with ``My chats with bots like ChatGPT will stay private,'' and \textit{agreed} they were concerned about personal information being misused by AI. Together, our findings reflect consistent distrust in the management and collection of personal information, including with LLM-based chatbots.
Our survey data indicate users are unwilling to provide chatbots access to emails, search history, or devices in exchange for better services—including premium-tier (\$200/month) subscriptions. Similarly, participants were unwilling to have chatbot data used to improve their online experience, especially on social media. These results depict general skepticism about chatbot data integration relevant to AI companies and regulators. Agentic AI systems promise to execute complex actions on behalf of users \cite{acharya2025agentic,kapoor2024ai}, and while early demonstrations are technically impressive, they depend upon extensive sensitive user data that users seem unwilling to trade. We establish a straightforward starting point for investigating consumer tradeoffs between privacy and advanced AI by showing \$200.00 in value is insufficient for unrestricted agentic data access.

Such unwillingness makes sense if adherence to privacy norms is both scarce and valuable in an online economy driven by user data \cite{lammi2019data}. In contextual integrity, the three significant factors isolated using our LMMs (informed consent, anonymity, and privacy) align with ``transmission principles,'' or aspects of information exchange governed by expectations and rights of data subjects \cite{malkin2022contextual}. This suggests that for emerging technology like LLM-based chatbots, users judge appropriateness based on well-understood governing principles. Our participants signaled the most meaningful aspect when considering responsible data use was simply being asked in advance (consent) and having personal information removed.

Our focus on transmission principles may reflect a post-hoc truth (\textit{i.e. that privacy and consent matter)}, but our results also point to emerging areas of interest to ethics researchers and policymakers \cite{susser2019notice}. Respondents described chatbot data as sensitive and expressed doubt it would remain private, and large pluralities report regularly discussing sensitive topics like personal finances, legal guidance, and health with chatbots. The ethical use of emerging technologies is more complicated than pointing to a privacy paradox \cite{martin2016measuring, dienlin2015privacy}, but growing work demonstrates users are unaware of how much sensitive data they exchange with chatbots, how this might impact their lives if disclosed, and ramifications for having chat data seized or used adversarially by law enforcement. Moreover, while respondents value informed consent, inconsistencies between privacy concerns and disclosure behaviors, combined with chatbots' complexity and opacity, suggest meaningful consent is challenging to achieve \cite{winograd2022loose, atata2024ai}. Consequently, traditional notice-and-consent paradigms, shown ineffective in protecting online privacy \cite{barocas2009notice, hijjawi2024end}, may be inadequate for addressing chatbot privacy concerns. We believe our results establish a significant baseline about factors that matter to end-user privacy notions and value users assign to chatbot interactions. We summarize findings in Table \ref{tab:implications}.

\subsection{Limitations}
One of the central limitations of our approach is that we restricted our participant population to people residing in the U.S. This was primarily due to financial constraints, and we are in the process of expanding this work to a comparative study of AI and chatbot privacy norms around the world. Moreover, as described in the Appendix, we deviated from our preregistration, most notably by adding our questions about personalization, privacy valuation, and integration of chat data. We thus note that results from these new questions should \textit{not} be viewed as having passed the ``severe'' test of preregistration. However, because we were also \textit{more} conservative with our data analysis than declared, future work might examine non-significant but marginal effects identified by our study, especially for Content and Location, as we believe we are more likely to make Type 2 errors, rejecting true hypotheses. Finally, future work might further develop our findings within the contextual integrity framework. Though we situate our results within the theory of contextual integrity, our methods cannot isolate more complex information flows without increasing the chances of making Type 1 errors.
\section{Conclusion}

We set out to understand the emerging privacy norms around chatbots. Our work demonstrated both broad concern about the use of chatbot data, particularly when data is integrated across contexts, as well as the primacy of informed consent, anonymity, and privacy (associated with transmission principle in contextual integrity) for determining the appropriateness of information flows with LLM-based chatbots.
\section*{Researcher Positionality}

This research highlights public concerns surrounding the privacy risks attendant to a novel emerging technology, LLM-based chatbots. Given the multidisciplinary nature of our work, we sought to include the expertise of scholars from many disciplines on this research, including those who characterize their primary research domains as social science, computer science, statistics, and public policy. In addition to their intellectual diversity, the authors of this work are also demographically diverse, reflective of a wide array of racial, ethnic, gender, and national identities. We believe that these forms of diversity ultimately benefited the paper. 

\section*{Adverse Impacts}

We do not anticipate immediate harms from the findings of this paper, as we focus our attention on the privacy concerns of Americans, and hope that our work will be used to foster more responsible design and public policy with respect to LLM-based chatbots. By focusing on Americans, we acknowledge that we exclude many important perspectives; however, as noted in the Limitations section, we expect to soon expand this work to several locations around the world.

\section*{Ethical Considerations}

Theories of contextual integrity have long sought to bridge the divide between social science and ethical theory as it pertains to the modern information economy. While the results of this work are primarily empirical and belong to the domain of social science, our research contributes perspectives that can inform normative ethical work with respect to the responsible handling of data produced in interaction with AI chatbots, assistants, and agents.
\section{Data Availability Statement}
\noindent
\begin{minipage}{\columnwidth}
\raggedright
The data and software developed for this study have been archived and are openly available in Dataverse: \url{https://doi.org/10.7910/DVN/M6ABJ3} and Github \url{https://github.com/WeberLab-UW/chatbot-privacy}
\end{minipage}
\section{Acknowledgments}
This research was supported by grants from the Sloan Foundation (G-2018-11217) and the Institute for Museum and Library Services (RE-252290-OLS-22). 

\bibliography{references}

\section{Appendix}

\section{Pilots}

To validate our survey instrument and set the level of compensation for our study, we conducted one informal, internal pilot followed by four formal pilots of our survey on the CloudResearch Connect platform \cite{hartman2023introducing}. In the internal pilot, we asked four graduate students at our University about which of several formats (plain text, grid-based, grid and plain text) made our factorial vignettes easiest to reason about. The students uniformly preferred the plain text format, which was also used previously by \citet{gilbert2021measuring}. Thus, in our first formal pilot, we A/B tested the full survey with two different versions of the plain text format for the vignettes: one with annotation identifying the factors in each vignette, and one presenting vignettes without annotation. $N$=25 participants saw each version, and independent samples $t$-tests demonstrated a marginal but non-significant effect showing that participants who received the unannotated version viewed the vignettes as easier to understand and complete than participants who received the annotated version. We also observed that participants were able to complete the unannotated version slightly more quickly, though this difference was also marginal and non-significant. Based on these marginal and nonsignificant but nonetheless directionally consistent results, we chose to use the unannotated plain text version of the factorial vignette for our study, and we used only this version in our subsequent pilots.

The remaining formal pilots sought to ensure data quality and determine compensation. In the second pilot, we added an attention check requiring that participants spend at least five seconds on each vignette. Surprisingly, this notably increased the amount of time needed by participants to complete the survey, and lowered participants' approval of the survey. In the remaining pilots, we validated that removing this new attention check restored participant approval of the survey (as measured using CloudResearch Connect's survey of our respondents), and we instead implemented a simple time log allowing us to review survey data for outliers. We also increased the compensation for the survey from \$3.00 to \$4.00; while internal pilots had suggested that participants would need 10-15 minutes to complete the survey, formal pilots suggested most participants would need 16-24 minutes. Note that each formal pilot recruited $N$=25 participants.

\section{Preregistered Study}

We submitted a pre-registration of our hypotheses and planned analysis with the Open Science Foundation in April 2025. Our goal in preregistering this study was to provide readers with a transparent description of our thinking as generated hypotheses, collected data, and analyzed and reported findings. We believe that the hypotheses and data analysis described in this paper would be judged to be thoroughly consistent with our preregistration, and that our data analysis would be considered more conservative than what we initially registered, by virtue of using both a larger model allowing for more robust mechanisms for correction, as well as a conservative approach to adjusting for the heteroscedasticity of residuals. For full transparency, the original pre-registered study is reported here: \url{https://osf.io/f43tb}. In accordance with OSF best practices, we documented modifications made to the orginal pre-registered study, including when the modifications where made, in our Statement of Transparent Changes: \url{https://osf.io/xs5cm}. 

\section{Interaction Effects}

As noted in the Results section of the paper, though we hypothesized that responses to the Privacy Attitudes questions indicating greater privacy concern would interact with the Consent/Awareness factor, such that we would observe increases in the level of concern when informed consent was not obtained, we did not observe any such interaction, as described in full in Table \ref{tab:interactions-consent}.

On the other hand, as shown in Table \ref{tab:interactions-locations}, we did observe a significant interaction in the Concern model for responses to the ``I am concerned that online companies are collecting too much personal information'' and both the European Union level and the China level of the Location factor, with $\beta$ values indicating increased appropriateness. On the other hand, we observe a reduction in appropriateness for the China level when computing an interaction with the ``In general, I believe that my conversations with chatbots like ChatGPT will remain private'' privacy attitudes question. We had hypothesized that participants who expressed greater concern about privacy via the attitudes questions would be more likely to deem information flows to a Receiver in the United States (the country in which all participants reside) as appropriate and information flows to a Receiver in China as less appropriate and a greater cause for concern. The only effect consistent with that hypothesis is the interaction between the China level and the conversations with chatbots will remain private question. However, because this effect seems to run counter to the interaction between the China level and the concern about online companies collecting too much information question, as observed in the opposing direction of the $\beta$ values, we cannot reliably draw any conclusion about this result, and it is not consistent with our hypothesis.

\clearpage

\begin{table*}[t]
\centering
\small
\begin{tabular}{p{7.5cm}|rrrr|rrrr}
\toprule
\multicolumn{1}{c|}{\textbf{Consent x Privacy Measure}} & \multicolumn{4}{c|}{\textbf{Appropriateness}} & \multicolumn{4}{c}{\textbf{Concern}} \\
\midrule
\textbf{} \ & Est & \textit{SE} & \textit{t} & \textit{p} & Est & \textit{SE} & \textit{t} & \textit{p} \\
\midrule
'In general, I believe privacy is important.' & -0.35 & 0.08 & -4.58 & $<$0.01 & 0.46 & 0.08 & 5.88 & $<$0.01 \\
\hspace{1em}x not informed that your data was collected & 0.06 & 0.05 & 1.5 & 0.26 & 0.04 & 0.04 & 1.10 & 0.27\\
\hspace{1em}x informed that your data was collected & 0.02 & 0.04 & 0.56 & 0.58 & -2.29 & 4.37 & -0.52 & 0.60\\
\midrule
‘In general, I trust websites.' & 0.22 & 0.04 & 5.41 & $<$0.01 & -0.19 & 0.04 & -4.78 & $<$0.01\\
\hspace{1em}x not informed that your data was collected & -0.02 & 0.03 & -0.53 & 0.59 & 0.01 & 0.05 & 0.29 & 0.77 \\
\hspace{1em}x informed that your data was collected & 0.03 & 0.03 & 1.12 & 0.26 & -0.04 & 0.05 & -0.97 & 0.34\\
\midrule
'I am concerned that online companies are collection too much personal information.' & -0.27 & 0.05 & -5.57 & $<$0.01 & 0.31 & 0.05 & 6.32 & $<$0.01\\
\hspace{1em}x not informed that your data was collected & 0.06 & 0.03 & 1.83 & 0.07 & 0.01 & 0.03 & 0.46 & 0.65 \\
\hspace{1em}x informed that your data was collected & -0.02 & 0.03 & -0.67 & 0.50 & -0.09 & 0.03 & 0.80 & 0.42 \\
\midrule
'I have privacy concerns about my conversations with ChatGPT.' & -0.19 & 0.04 & -4.41 & $<$0.01 & 0.29 & 0.04 & 6.84 & $<$0.01 \\
\hspace{1em}x not informed that your data was collected & 0.03 & 0.03 & 1.30 & 0.20 & -0.03 & 0.02 & -1.46 & 0.14 \\
\hspace{1em}x informed that your data was collected & -0.02 & 0.03 & -0.72 & 0.47 & -0.01 & 0.03 & -0.38 & 0.71\\
\midrule
‘In general, I believe that my conversations with chatbots like ChatGPT will remain private.' & 0.15 & 0.04 & 3.81 & $<$0.01 & -0.18 & 0.04 & -4.95 & $<$0.01 \\
\hspace{1em}x not informed that your data was collected & 0.03 & 0.03 & 1.01 & 0.31 & -0.00185 & 0.03 & -0.06 & 0.94\\
\hspace{1em}x informed that your data was collected &  0.01 & 0.03 & 0.50 & 0.62 & 0.03 & 0.03 & 1.20 & 0.23\\
\bottomrule
\end{tabular}
\caption{We observed no significant interaction effects between Consent and responses to the five privacy attitudes questions.}
\label{tab:interactions-consent}
\end{table*}

\begin{table*}[t]
\centering
\small
\begin{tabular}{p{7.5cm}|rrrr|rrrr}
\toprule
\multicolumn{1}{c|}{\textbf{Location x Privacy Measure}} & \multicolumn{4}{c|}{\textbf{Appropriateness}} & \multicolumn{4}{c}{\textbf{Concern}} \\
\midrule
\textbf{} \ & Est & \textit{SE} & \textit{t} & \textit{p} & Est & \textit{SE} & \textit{t} & \textit{p} \\
\midrule
'In general, I believe privacy is important.' & -0.31 & 0.07 & -4.37 & $<$0.01 & 0.43 & 0.07 & 6.10 & $<$0.01 \\
\hspace{1em}x China & -0.04 & 0.05 & -0.76 & 0.45 & 0.04 & 0.04 & 1.10 & 0.27\\
\hspace{1em}x The European Union & -0.02 & 0.05 & -0.35 & 0.72 & -2.29 & 4.37 & -0.52 & 0.60\\
\midrule
‘In general, I trust websites.' & 0.25 & 0.04 & 6.39 & $<$0.01 & -0.19 & 0.04 & -4.78 & $<$0.01\\
\hspace{1em}x China & -0.04 & 0.03 & -1.18 & 0.23 & 0.04 & 0.03 & 1.08 & 0.28 \\
\hspace{1em}x The European Union & -0.03 & 0.03 & -1.06 & 0.29 & 0.05 & 0.03 & 1.61 & 0.11 \\
\midrule
'I am concerned that online companies are collecting too much personal information.' & -0.30 & 0.05 & -6.38 & $<$0.01 & 0.37 & 0.05 & 7.68 & $<$0.01\\
\hspace{1em}x \cellcolor{gray!30}China & \cellcolor{gray!30}0.06 & \cellcolor{gray!30}0.03 & \cellcolor{gray!30}2.05 & \cellcolor{gray!30}$<$\textbf{0.05} & -0.04 & 0.03 & -1.37 & 0.17\\
\hspace{1em}x \cellcolor{gray!30} The European Union & \cellcolor{gray!30}0.06 & \cellcolor{gray!30}0.03 & \cellcolor{gray!30}2.01 & \cellcolor{gray!30}$<$\textbf{0.05} & \cellcolor{gray!30} -0.09 & \cellcolor{gray!30} 0.03 & \cellcolor{gray!30} -3.18 & \cellcolor{gray!30} $<$\textbf{0.01}\\
\midrule
'I have privacy concerns about my conversations with ChatGPT.' & -0.21 & 0.04 & -5.16 & $<$0.01 & 0.29 & 0.04 & 7.26 & $<$0.01 \\
\hspace{1em}x China & 0.04 & 0.03 & 1.61 & 0.11 & -0.01 & 0.03 & -0.29 & 0.77 \\
\hspace{1em}x The European Union & 0.04 & 0.03 & 1.35 & 0.18 &  -0.05 &  0.03 &  -1.97 & 0.05\\
\midrule
‘In general, I believe that my conversations with chatbots like ChatGPT will remain private.' & 0.20 & 0.04 & 5.37 & $<$0.01 & -0.18 & 0.04 & -5.02 & $<$0.01 \\
\hspace{1em}x \cellcolor{gray!30}China & \cellcolor{gray!30}-0.06 & \cellcolor{gray!30}0.03 & \cellcolor{gray!30}-2.00 & \cellcolor{gray!30}$<$\textbf{0.05} & 0.01 & 0.03 & 0.42 & 0.67\\
\hspace{1em}x The European Union & -0.05 & 0.03 & -1.80 & 0.07 & 0.03 & 0.03 & 1.01 & 0.31\\
\bottomrule
\end{tabular}
\caption{Though we observe some significant interactions between the factor Location and the privacy attitudes questions, the interactions are directionally inconsistent, particularly with respect to the China level, rendering them difficult to interpret.}
\label{tab:interactions-locations}
\end{table*}

\end{document}